\documentclass[11pt]{article}

\usepackage[final]{acl}
\usepackage{subcaption}    %

\usepackage{times}
\usepackage{latexsym}

\usepackage[T1]{fontenc}

\usepackage[utf8]{inputenc}

\usepackage{microtype}

\usepackage{inconsolata}

\usepackage{graphicx}

\usepackage{amsmath}
\usepackage{amssymb}
\usepackage{mathtools}
\usepackage{amsthm}
\usepackage{booktabs}

\usepackage{multirow}

\title{Embedding Inference Attack}

 \author{Cedric Fitiavana Raelijohn \and Sébastien Gambs \and Jean-Francois Rajotte\\
         Université du Québec à Montréal, Canada\\
         \texttt{raelijohn.cedric\_fitiavana@courrier.uqam.ca}, \texttt{gambs.sebastien.@uqam.ca},\\ \texttt{rajotte.jean-francois@uqam.ca}\\
         }

\begin{document}
\maketitle
\begin{abstract}
Embedding models are essential components of modern Information Retrieval (IR) systems, yet they are typically hidden behind APIs. 
Recent works have shown that dense IR system can lead to security vulnerabilities such as embedding inversion attacks. 
However, such attacks usually require that the attacker knows the embedding model for the attack to be applicable.
In this paper, we study IR systems under a black-box setting in which the adversary observes only the unordered set of retrieved documents, without ranking or similarity scores. 
We demonstrate that in such contexts, tailored queries allow an adversary to identify which embedding model is in use from a set of known model candidate, which we coin as an embedding inference attack (EIA). 
We also show that certain queries remain discriminative even when the system includes a reranker as a potential defense mechanism. 
We further validate our method on a real Retrieval-Augmented Generation (RAG) system, in which the tailored queries bypass the LLM’s tendency to reject inputs it does not recognize as well-formed questions. 
Finally, we propose and evaluate other mitigation strategies such as similarity thresholds.
\end{abstract}

\section{Introduction}

With the growth of data and the rise of artificial intelligence (AI), IR systems have evolved into modern applications, such as dense (or semantic) IR, in which documents and queries are encoded into vector representations (embeddings vectors) using a neural network, denoted as embedding models~\cite{karpukhin-etal-2020-dense}.
This approach enables semantic matching rather than exact keyword matching, significantly improving retrieval performance, especially in large scale datasets.
More recently, Retrieval-Augmented Generation (RAG) has emerged as a popular paradigm combining dense retrieval with general-purpose large language models (LLMs) to generate context-aware responses. 
This is especially relevant for domain-specific fields such as finance, healthcare and education~\cite{ragusage1,ragusage2}.

With respect to privacy, embedding vectors are often assumed to be safe to expose because they are only numerical values that do not directly convey an interpretable information for a human. 
However, recent works have demonstrated the feasibility of conducting embedding inversion attacks, which allow to reconstruct the text represented by embeddings vectors~\cite{vec2text,zsinvert}.
For instance, this could happen in the common situation in which a RAG system is based on publicly accessible data and is implemented to help users navigate a large set of documents, such as documentation, policies, etc.
In this situation, the queries could contain sensitive information such as personal medical conditions.
Embedding inversion attacks often rely on strong assumptions about the adversary's capabilities such as access to large sets of leaked text–embedding pairs or knowledge of the victim’s embedding model and architecture \cite{vec2text,teia,zsinvert}.
We show here how IR systems based on publicly available datasets relaxes these assumptions, making them more vulnerable to inversion attacks.

In this work, we introduce the concept of Embedding Inference Attack (EIA), which aims to infer the embedding model used by the target system in a black-box setting in which the adversary has only access to the API of the system. 
This setting is realistic because embedding models are commonly accessible only through API, for instance in Semantic Scholar AI-powered research tool and OpenEvidence searching PubMed and medical journals.
As a specific use case, we are particularly interested in IR for education in which high school students can ask questions about their homework but might inadvertently leak personal information (\emph{e.g.}, name, age or medical conditions of some family members).
Knowledge of the embedding model in an IR system such as a RAG can also unlock security vulnerabilities other than inversion attacks, such as corpus poisoning~\cite{Zhong2023PoisoningRC}, membership inference in a vector database~\cite{miavdb2025} and search-engine optimization (SEO) attacks \cite{BenTov2024GASLITEingTR}.%
To the best of our knowledge, our work is the first black-box attack aiming to infer the embedding model on IR systems.

\paragraph{Contributions.} Our contributions can be summarized as follows.
\begin{itemize}
    \item We propose EIA, a novel approach to infer the embedding model of a black-box IR system. 
    We also show how to instantiate this attack through a set of tailored queries allowing an adversary to identify which embedding model is used from a set of known candidates.
    \item We experimentally validate our method on multiple open-weight embedding models and a Retrieval-Augmented Generation (RAG) system, in which the tailored queries bypass the LLM’s tendency to reject inputs that do not correspond to well-formed questions.
    \item We demonstrate that certain queries remain discriminative even when the system includes a reranker or uses similarity thresholds as possible mitigation strategies against an EIA.
    \item We showcase the performance of our attack on real-world question-answering (QA) datasets, including MS MARCO and an educational QA dataset collected from Alloprof.
\end{itemize}

\paragraph{Outline.} The outline of the paper is as follows.
First in Section~\ref{sec:background}, we provide the background notions on information retrieval and present the system and adversary models that we consider before describing the related work on attacks involving the embedding model in Section~\ref{sec:related}. 
Afterwards, we introduce our EIA approach and how to implement it by generating discriminative queries in Section~\ref{sec:eia}. 
Then, we describe the experimental setting in Section~\ref{sec:experiments} before reporting on the results in Section~\ref{sec:results}. 
Finally, we evaluate possible countermeasures in Section~\ref{sec:defense} before concluding in Section~\ref{sec:conclusion} and then discussing the limitations.

\section{Background}
\label{sec:background}
In this section, we first provide an overview of information retrieval setting considered before describing the systems and adversary models.

\subsection{Overview of information retrieval}
\paragraph{Embedding model and information retrieval.}
In a dense IR system~\cite{karpukhin2020dense}, queries and documents are converted by an embedding model into dense embedding vectors.
A desired characteristic for IR is that relevant documents should to be close to a given query in the corresponding embedding space.
Conversely, irrelevant documents to the query should be located farther away.
To retrieve the corresponding documents for a given query, similarity measures between their corresponding vectors and the query are computed.
Classical similarity measures include the Euclidean distance or cosine similarity, the latter being the most commonly used~\cite{vectordatabases}.

\paragraph{Reranker.}
A reranker is generally a model that is used to refine and re-order the initial set of retrieved documents to achieve better relevance for a given query \cite{re3qa}.
The retrieval pipeline with a reranker generally operates in two phases.
First, an initial retrieval step performed typically through embedding models returns a relatively large set of documents (\emph{e.g.}, top-50).
Second, these documents, along with the query, are then passed to the reranker, which selects the top-$k$ most relevant documents (\emph{e.g.}, top-3 or top-5)~\cite{calam2023rerankers}.

Rerankers primarily use cross-encoder architectures, in which the query and each document are concatenated and processed together through a transformer model to produce a relevance score~\cite{calam2023rerankers,lu2025crossencoder}. 
This contrasts with the embedding models used in the initial retrieval, which encode queries and documents independently and compute similarity via a distance metric~\cite{reimers2019sentencebert}. 
The cross-encoder approach enables deeper semantic understanding of the query-document relationship obtained from the first retrieval stage, as it processes each individual query-document pair (\emph{i.e.}, the query paired with each of the top-$k$ retrieved documents) resulting in a more expensive computation, thus explaining the latency tradeoff~\cite{calam2023rerankers,pandit2025reranking}.
Recently, researchers have demonstrated that large language models (LLMs), such as ChatGPT and Mistral variants fine-tuned for ranking tasks, can also serve as rerankers~\cite{pandit2025reranking}.
Rather than directly generating answers, these models evaluate the relevance of candidate documents with respect to a query and can leverage their stronger reasoning capabilities to improve ranking quality~\cite{rankllm}.

\subsection{System and adversary models}

We consider a black-box information retrieval system that is accessible only through an API.
In this system, a query, given as a sequence of tokens $q \in \mathbb{V}^n$, is encoded by an embedding model
$\phi : \mathbb{V}^n \rightarrow \mathbb{R}^d$ into a fixed-length query embedding vector
$e \in \mathbb{R}^d$.
Similarly, the same embedding model $\phi$ converts a document $D \in \mathbb{V}^n$, also represented as a sequence of tokens,
into a fixed-length document embedding vector
$f = \phi(D) \in \mathbb{R}^d$.

The system ranks documents by computing the similarity between the query embedding $e_i$ and document
embeddings $f_j$, and returns the top-$k$ most relevant documents~\cite{zeng-etal-2024-good}.
Several similarity  metrics are commonly used to compare vector representations in information retrieval system. 
Cosine similarity is independent of vector magnitude and calculates the cosine of the angle between two vectors. 
Its value ranges from -1 to 1, with smaller angles corresponding to higher similarity. 
Euclidean (L2) distance, in contrast, calculates the straight-line distance between two points in vector space, in which smaller distances indicate greater similarity. 
Dot product (inner product) similarity is calculated as the product of the two vectors' magnitudes and the cosine of the angle between them.\cite{Manningetal2008}
In this work, we use cosine similarity exclusively, as its insensitivity to vector magnitude is critical when comparing high-dimensional text embeddings of varying length.

The objective of the attacker is to infer the embedding model $\phi_v$ used by the IR system. 
In this context, the attacker is following the black-box adversary setting in which he can only query the victim system with any questions and only has access to the (unranked) retrieved documents.
Moreover, similarly to most of the work in the literature we make the following assumptions:
\begin{enumerate}

\item The retrieval system uses a publicly available source of documents (\emph{e.g.}, Wikipedia) and for a given query $q_i$, it returns a set $S_i$ of $k$ unranked documents, which are outputs that can differ only by the order of documents. 
For instance, the set $d_1, d2, d3$ is considered to be equivalent to $d_3,d_1,d_2$~\cite{manning2008evaluation}.
\item One of the candidate model is the victim model. 
This assumption is realistic because several retrieval-system vendors publicly disclose the embedding models available through their services. 
For example, OpenAI and Cohere both document their supported embedding models and associated API identifiers \cite{OpenAI2024Embedding,Cohere2026Embed}.
Thus, in these situations the possible set of candidate embedding models can be considered to be public information.
\end{enumerate}
However in contrast to other previous works, \emph{we consider the challenging setting in which the attacker does not have access to the embedding vectors}.

\section{Related work}
\label{sec:related}

\paragraph{Model stealing.}
\citet{dziedzic2023sentence} have shown that it is easy to steal sentence embedding encoders via a black-box API that returns only the vector of the input text and might also expose metadata including the architecture family (\emph{e.g.}, BERT or RoBERTa) and possibly pre-training datasets. 
Afterwards, the adversary can instantiate a publicly available pre-trained transformer checkpoint with the same architecture or at least the same output dimensionality and train it with a dataset of sentence pairs drawn from any open-source text distribution.
Another work by~\citet{tamber-etal-2025-cant} demonstrates that API-only access can be enough to efficiently replicate the effectiveness of commercial embedding models by distilling them into local BERT-based models while preserving most retrieval performance. 
In their work, the attacker only sees input text and output embedding vectors and has unrestricted access to the embedding API.

\paragraph{Model inference and similarity.}
\citet{Pasquini2024LLMmapFF} have shown how to send very few crafted queries to identify specific LLMs based on their answers. 
In this context, the attacker can send queries to a victim RAG application and observe the outputs. 
These outputs are then fed to a separate ML-based inference model predicting which of 42 candidates LLM versions is generating the answers, achieving over 95\% accuracy.
\citet{beyondbench} compares embedding models with various similarity measures in order to help selecting the most suitable for a given IR or RAG system.
The authors determine pair-wise model similarities based on embedding vectors with Central Kernel Alignments (CKA) or resulting document retrieval.
The models' document retrievals are compared with and without ranking.
Their experiments reveal clear intra‑ and inter‑family clusters when comparing retrieval over large top-$k$ which is essentially the full dataset corpus (3 600 document chunks), but show that at realistic RAG settings \emph{e.g.}, top‑10, retrieval similarity between models is low and highly variable, implying that models with similar representations can still surface largely distinct document sets.

\paragraph{Inversion attacks.}
\citet{vec2text} introduce Vec2Text, an embedding inversion attack to reconstruct text produced by embedding models such as OpenAI Text-embedding-ada-002 and gtr-t5-base. 
In this work, the attacker is assumed to have a large dataset of leaked text–embedding pairs from the target encoder or from another leaked dataset, in which case the attacker needs to train the decoder so that it can generate candidate texts whose embeddings are close to a given target embedding.
Vec2Text frames the inversion attack as a controlled generation in which starting from an initial hypothesis, the model iteratively corrects the text by repeatedly re‑embedding the current guess with black‑box access to the encoder and conditioning on the target embedding, the current embedding and their difference to move the hypothesis in embedding space towards the target.

\citet{geia} have propose GEIA for Generative Embedding Inversion Attack, which uses a conditional language model training to inverse sentence embeddings directly. 
During the training, the decoder is trained by doing teacher forcing, which means that at each step the decoder is fed with the ground‑truth of the previous token rather than its own predicted token to learn to reconstruct the original sentence from its embedding.
Here also, the attacker is assumed to have an auxiliary dataset similar to the distribution of the dataset of the victim and a black-box access in which he can only query the model victim to obtain an text-embedding pairs from his corpus. 

\citet{teia} have introduced TEIA for Transferable Embedding Inversion Attack, which is an inversion attack that learns a surrogate model of the embedding model victim. 
The setting is more restricted than other works in the sense that the attacker cannot query or access the embedding model victim but needs to have access to a small leaked set of text-embeddings pairs from it. 
\citet{teia} re-implemented the GEIA attack as a ``Direct Attack'' baseline and evaluate both methods on the same datasets and the same embeddings. 
Across all reported dataset–encoder combinations, TEIA consistently achieves higher reconstruction quality than GEIA in terms of Rouge-L\footnote{Rouge-L is a metric used to evaluate how well the generated summary preserves the ordering and content of the reference~\cite{lin-2004-rouge}.}
and embedding cosine similarity.
For example, on QNLI dataset with OpenAI embedding, RougeL improves from 0.1433 to 0.2226 and cosine similarity from 0.2797 to 0.4772. 
Similarly on AGNEWS dataset, TEIA increases RougeL from 0.0612 to 0.1271 and cosine similarity from 0.1162 to 0.4301.

\citet{zsinvert} have introduced ZSinvert, a Universal Zero-shot Embedding Inversion attack that does not require prior decoder training but the attacker is assumed to have access to the exact embedding vector and black‑box query access to the encoder that produced it. 
ZSinvert inverts an embedding vector by using an LLM with a modified beam search that directly optimizes cosine similarity between the encoder’s output on candidate texts and the target embedding.
More precisely, it first generates a rough candidate with an open‑ended prompt, then repeatedly paraphrases and corrects this candidate while always being guided by the target embedding without ever training an encoder‑specific inversion model.

\citet{algen} have proposed a few-shot inversion attacks called ALGEN for Alignment and Generation, which aligns the victim embeddings to the attack space and using a generative model to reconstruct the original text. 
The attacker is assumed to have a small number of leaked embedding vectors (without the corresponding original documents) and has black‑box access to the victim encoder.
The attacker independently trains a separate ``local'' embedding-to-text generator on public data using their own model.
Afterwards, a linear transformation matrix is learned that maps embeddings from the victim’s embedding space to the attacker’s embedding space.

Finally, \citet{zero2text} have introduced Zero2Text, a training‑free cross‑domain embedding inversion. 
In this setting, the attacker does not need any prior leaked embedding-text pairs from the victim and is assumed to have only a black-box access in which he can query the victim embedding model via its API and see its output.
Zero2Text uses a pretrained LLM to generate candidate sentences and a dynamically updated linear projection (learned via online ridge regression) to map the attacker’s local embedding space into the victim’s embedding space for scoring and selection.

\begin{table*}[h!]
\centering
\small
\setlength{\tabcolsep}{4pt}
\begin{tabular}{l c c p{3cm} p{5.5cm}}
\toprule
\textbf{Attack} &
\textbf{Encoder access} &
\textbf{Training} &
\textbf{Data knowledge} &
\textbf{Core Idea} \\
\midrule

Vec2Text~\cite{vec2text} &
Black-box &
$\checkmark$ &
Large corpus leak of text--embedding pairs &
Train an inversion model and iteratively refine candidate texts by minimizing embedding differences. \\
\addlinespace

GEIA~\cite{geia} &
Black-box &
$\checkmark$ &
A similar distribution corpus and queried text--embedding pairs &
Train a conditional language model to directly reconstruct text from embeddings. \\
\addlinespace

TEIA~\cite{teia} &
None &
$\checkmark$ &
Small leaked set of text--embedding pairs &
Learn a surrogate encoder and perform inversion through transferability. \\
\addlinespace

ZSInvert~\cite{zsinvert} &
Black-box &
$\times$ &
Target embedding only &
Use LLM-guided beam search with adversarial decoding to optimize embedding similarity without training. \\
\addlinespace

ALGEN~\cite{algen} &
Black-box &
$\checkmark$ &
Small set of leaked victim embeddings only &
Align embedding spaces and reconstruct text using a locally trained generator. \\
\addlinespace

Zero2Text~\cite{zero2text} &
Black-box &
$\times$ &
No leakage &
Use LLM and online embedding-space projection without training a specific decoder. \\
\bottomrule
\end{tabular}
\caption{Attacker assumptions in existing embedding inversion attacks.}
\label{tab:inversion_comparison}
\end{table*}

Table~\ref{tab:inversion_comparison} summarizes embedding reconstruction attacks according to the attacker capabilities and the underlying attack approach.
All attacks require either access to the embedding model or a set of leaked text-embedding pairs.

\section{Embedding inference attack description}
\label{sec:eia}

For a given query $q_i$, the retrieved documents from the victim $S^v_i$ and candidate $S^c_i$ are compared using a Jaccard index, denoted by the following equation:
\begin{equation}
J_i= \frac{|S^v_i \cap S^c_i|}{|S^v_i \cup S^c_i|}.
\end{equation}
To perform the EIA, the attacker iteratively queries the retrieval system and compares the returned document sets. 
Two embedding models, denoted as $\phi_c$ (candidate) and $\phi_v$ (victim), are considered likely to be the same if the Jaccard index is maximized for any question, $J_i = 1 , \forall i$. 
However, even if $\phi_c \neq \phi_v$, it is possible for the Jaccard index to reach 1 for some queries.
This approach trivially extends to a set of model candidates $\{ \phi^i_c \}$ by querying until all but one candidate are eliminated.

\paragraph{Brute-force generation of discriminative queries.}
The attacker can use an LLM to generate a set of questions, one for each document of the document corpus used by the victim system.
The generated questions are then used to query the victim model to perform the EIA.
To limit the number of API calls on the IR system, the attacker aims to use a question that eliminates all the candidate models.
To identify queries that effectively distinguish between candidate models, the attacker compares the retrieved document sets produced by each model. 
For a given query $q_i$, the similarity between two models is measured using the Jaccard index $J_i$ from equation (1). 
Queries for which $J_i \neq 1$ indicate discrepancies in retrieval are therefore considered potentially discriminative.

Let $\mathcal{M}$ denote the set of candidate models. 
For each query $q_i$, we define an elimination score as:
\begin{equation}
E(q_i) = |\{ m \in \mathcal{M} \;|\; J_i^{(v,m)} \neq 1 \}|,
\end{equation}
in which $J_i^{(v,m)}$ represents the Jaccard index between the victim model $v$ and a candidate model $m$.
The elimination score $E(q_i)$ corresponds to the number of candidate models whose retrieval differs from the victim model. 
Intuitively, higher values of $E(q_i)$ indicate stronger discriminative power.
The attacker then selects the subset of queries that maximizes discrimination across all candidate models. 
In particular, queries with $E(q_i) = |\mathcal{M}|$ are ideal, as they allow the attacker to eliminate all incorrect candidates while queries with lower elimination scores are less effective and are discarded directly. 
A limitation of this brute-force attack is the high number of queries required to generate and test to find a single fully discriminative query.
To address this, hereafter we show how to create queries that are discriminative in a more informed manner.

\paragraph{Generation of adversarial discriminative queries.}
\label{sec:adversarial_discriminative_queries}
We have explored three types of adversarial discriminative queries: 
\begin{enumerate}
\item Random Strings, denoted as \textbf{Rnd. Str.}, which consist of sequences of arbitrary characters, such as for example ``fsaduifhsdaifnjklsdan''. 
These queries allow to explore parts of the embedding space that are not covered by the training of the LLM.
\item Non-Queries, denoted as \textbf{Non-Q.}, are instructions rather than actual questions. 
An example of Non-Queries is ``Short answer only'' 
and the rationale behind them is that they allow to reach parts of the embedding space not related to documents. 
\item Probing Queries, denoted as \textbf{Prob. Q.}, are questions designed to obtain information about the victim system itself. 
Examples of probing queries include ``Who is your developer or creator?'' or  ``Are you an open-source model or a proprietary model?'' and their main objective is to investigate areas of the embedding that are related to the model.
\end{enumerate}

\section{Experimental setting}
\label{sec:experiments}

\subsection{Datasets}
We have used two datasets for our experiments. 
The first one is the MS MARCO passage development dataset that was developed by Microsoft to support research in machine reading comprehension and information retrieval~\cite{Bajaj2016Msmarco}. 
This dataset is built from anonymized real user queries issued to the Bing search engine, paired with passages extracted from web documents. 
This subset contains 7000 queries, 8.8 million passages and 7400 relevance judgments (qrels) that was human-annotated, indicating which documents are considered relevant to each query. 
We further selected only the documents having associated qrels, which leads to a subset of 7400 documents.

The second one is an educational question–answering dataset, created to support the development and evaluation of question-answering and information retrieval systems in a real-world educational context~\cite{alloprof}.
It contains questions from K-12 students across different subjects including mathematics, science and geography along with explanatory answers provided by educators (which we denote as documents).
This dataset is primarily in French, as it is designed for French-speaking students in Québec but a subset includes English queries and documents.
We then restrict our experiments to this subset and selected 585 out of 30000 documents.

For both datasets, the documents are assumed to be publicly accessible, aligning with the assumptions of our adversary model in which no private data leakage is required.

\subsection{Information retrieval pipeline}

We evaluated our attack on a controlled IR pipeline hosted on a High Performance Computing cluster, using a single NVIDIA H100 GPU partitioned into two GPU of 40GB instances via NVIDIA Multi-Instance GPU (MIG) technology, along with 64 CPU cores and 64GB of RAM. 
The full evaluation required approximately 8 GPU hours without reranker and 12 -- 15 hours when using a reranker. 
The IR system uses an embedding model to encode both queries and documents before retrieving the top-$k$ documents based on cosine similarity. 

\paragraph{Embedding models.}
We have used 13 open-source embedding models that we grouped into five families.
The E5 models are BERT-based dual encoders trained with large-scale weakly supervised contrastive learning on query--document pairs for general-purpose text embeddings~\cite{wangee5}.
The English SBERT models (all-MiniLM-L6-v2, all-mpnet-base-v2) are Siamese BERT-style dual encoders from the Sentence-Transformers framework, optimized for sentence-level semantic similarity and retrieval~\cite{reimers2019sentencebert}.
The multilingual SBERT models, including LaBSE, employ multilingual Transformer dual encoders trained with translation ranking and contrastive objectives to yield language-agnostic sentence embeddings~\cite{reimers2019sentencebert,feng-etal-2022-language}.
gtr-t5-base is a T5 encoder-based dual encoder from the Generalizable T5 Retriever family \cite{ni-etal-2022-large}, while Qwen3-Embedding-4B from the Qwen3 family is specifically designed for text embedding and ranking tasks \cite{qwen3embedding}.

All models are lightweight dual encoders or encoder-style architectures commonly used for dense retrieval and semantic similarity, with embedding dimensionalities ranging from 384 to 2560. 
These models were selected based on their wide use in prior work and their suitability for retrieval-oriented evaluation, notably~\citet{brown_systematic_2025}. 
Table~\ref{tab:embedding_models} summarizes the models and their embedding dimensionalities.

\begin{table}[h!]
\centering
\resizebox{\columnwidth}{!}{%
\begin{tabular}{llc}
\toprule
\textbf{Family} & \textbf{Model} & \textbf{Dim.} \\
\midrule
\multirow{3}{*}{E5}
  & e5-small, e5-small-v2                        & 384  \\
  & e5-base, e5-base-v2                          & 768  \\
  & e5-large, e5-large-v2                        & 1024 \\
\midrule
\multirow{2}{*}{English SBERT}
  & all-MiniLM-L6-v2                             & 384  \\
  & all-mpnet-base-v2                            & 768  \\
\midrule
\multirow{3}{*}{Multilingual SBERT}
  & paraphrase-multilingual-MiniLM-L12-v2        & 384  \\
  & LaBSE                                        & 768  \\
  & paraphrase-multilingual-mpnet-base-v2        & 768  \\
\midrule
T5 encoder
  & gtr-t5-base                                  & 768  \\
\midrule
Qwen3
  & Qwen3-Embedding-4B                           & 2560 \\
\bottomrule
\end{tabular}%
}
\caption{Open-source embedding models evaluated in our experiments.}
\label{tab:embedding_models}
\end{table}

\paragraph{Retrieval parameters and query types.}
For each model, we retrieve the top-$k$ documents with: $k=3$ and $k=5$.
We evaluate four categories of query: generated questions obtained from Qwen2.5-1.5B-Instruct, Random Strings (Rnd. Str.), Non-Queries (Non-Q.) and Probing Queries (Prob. Q.) manually crafted by the attacker and described previously in Section~\ref{sec:adversarial_discriminative_queries}. 

\paragraph{Reranking.}
We evaluate our EIA performance with (+ reranker) and without reranker.
We used the \textit{cross-encoder/ms-marco-MiniLM-L6-v2} reranker model and use 25 documents as an initial retrieval.

\subsection{RAG evaluation}

To validate our approach in a more realistic deployment scenario, we conduct additional experiments using an open-source RAG system provided by AnythingLLM\footnote{\url{https://github.com/Mintplex-Labs/anything-llm}}. 
AnythingLLM is a free and open-source, all-in-one AI application in which user can host a LLM locally or build their RAG system without requiring technical coding background or expensive infrastructure~\cite{AnythingLLM2024}.
The RAG was setup by using three embedding models : all-MiniLM-L6-V2, nomic-embed-text-v1, e5-small as well as one large language model for answer generation : Qwen3 Vision 4B instruct. 
We have chosen $k = 3$ for document retrieval as well as an optional reranker.

For each of the embedding models, documents are stored on LanceDB, which is the default local vector database of AnythingLLM with a default tokenization of 1000 with an overlap of 20. 
AnythingLLM have different ways of interacting with the documents and the underlying LLM. 
More precisely, we evaluated our attack using two interaction modes.
In the \textbf{Chat mode}, the LLM combines both the embedded documents and its own general‑knowledge base to produce more conversational flexible responses while in the \textbf{Query mode} the system restricts its answers strictly to information retrieved from the embedded documents, treating each interaction as a focused question‑and‑answer session.

\subsection{Overall evaluation protocol}

We perform a round-robin evaluation over the embedding models, in which each model is treated in turn as the victim while the remaining models act as candidates.
For each query $q_i$, we compare the retrieved document sets between the victim system and candidate models using the Jaccard index as detailed previously. 
This allows us to quantify retrieval similarity and identify discriminative queries for model inference.
We report results across all configurations to analyze (1) the impact of $k$ on model distinguishability, (2) the impact of different query types as well as (3) the effect of reranking on the attack.
Table~\ref{table:configuration summary} summarizes of the different configurations used for the experimentation.
\begin{table}[htpb]
    \centering
    \begin{tabular}{lccc}
    \hline
    \textbf{Setup} & \textbf{\# Models} & \textbf{k}\\
    \hline
    IR Pipeline & 13 & 3, 5\\
    AnythingLLM & 3 & 3\\
    \hline
    \end{tabular}
    \caption{Summary of experimental configurations.}
    \label{table:configuration summary}
\end{table}

\section{Results}
\label{sec:results}

In this section, we first review the results of the EIA in the information retrieval pipeline before reporting the success of the attack on the RAG framework.

\subsection{Results on IR Pipeline}
\begin{figure}[h!]
\begin{subfigure}{0.23\textwidth}
\includegraphics[width=0.9\linewidth]{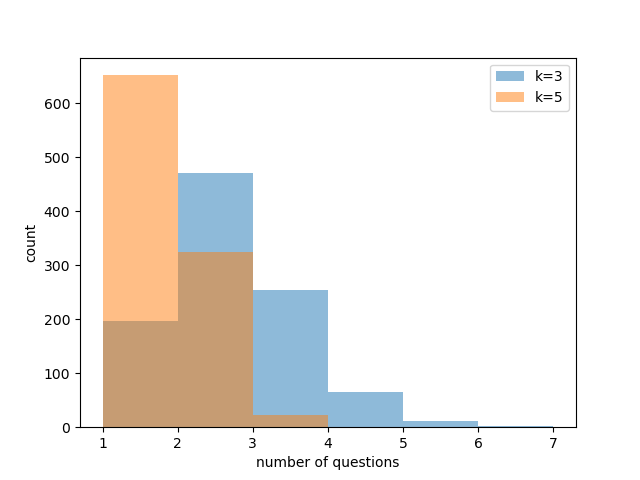} 
\caption{MSMarco}
\label{fig:ngen2disc_msmarco}
\end{subfigure}
\begin{subfigure}{0.23\textwidth}
\includegraphics[width=0.9\linewidth]{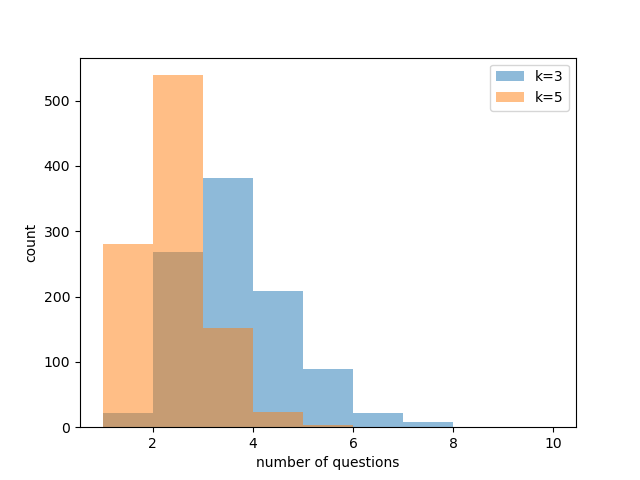}
\caption{Alloprof}
\label{fig:ngen2disc_alloprof}
\end{subfigure}
\caption{Distribution of the number of randomly generated questions needed to infer the embedding model.}
\label{fig:ngen2disc}
\end{figure}

\paragraph{Discriminative queries.}
As illustrated in Table~\ref{table:discriminative_query_summary_k3},  1249 generated queries are sufficient to successfully discriminate among all evaluated models when retrieving the top three documents ($k=3$).
Figure~\ref{fig:ngen2disc} shows the distributions of the number of randomly selected questions needed to successfully discriminate.
The histograms are produced by repeating the experiment of 1) selecting a document at random, 2) generating a question, 3) filtering out discriminated candidate embedding models and repeating until the victim model is the only candidate left.
On average, it usually takes few queries to discriminate the victim model from the candidate ones.

\begin{figure}[h!]
    \centering
    \includegraphics[width=\columnwidth]{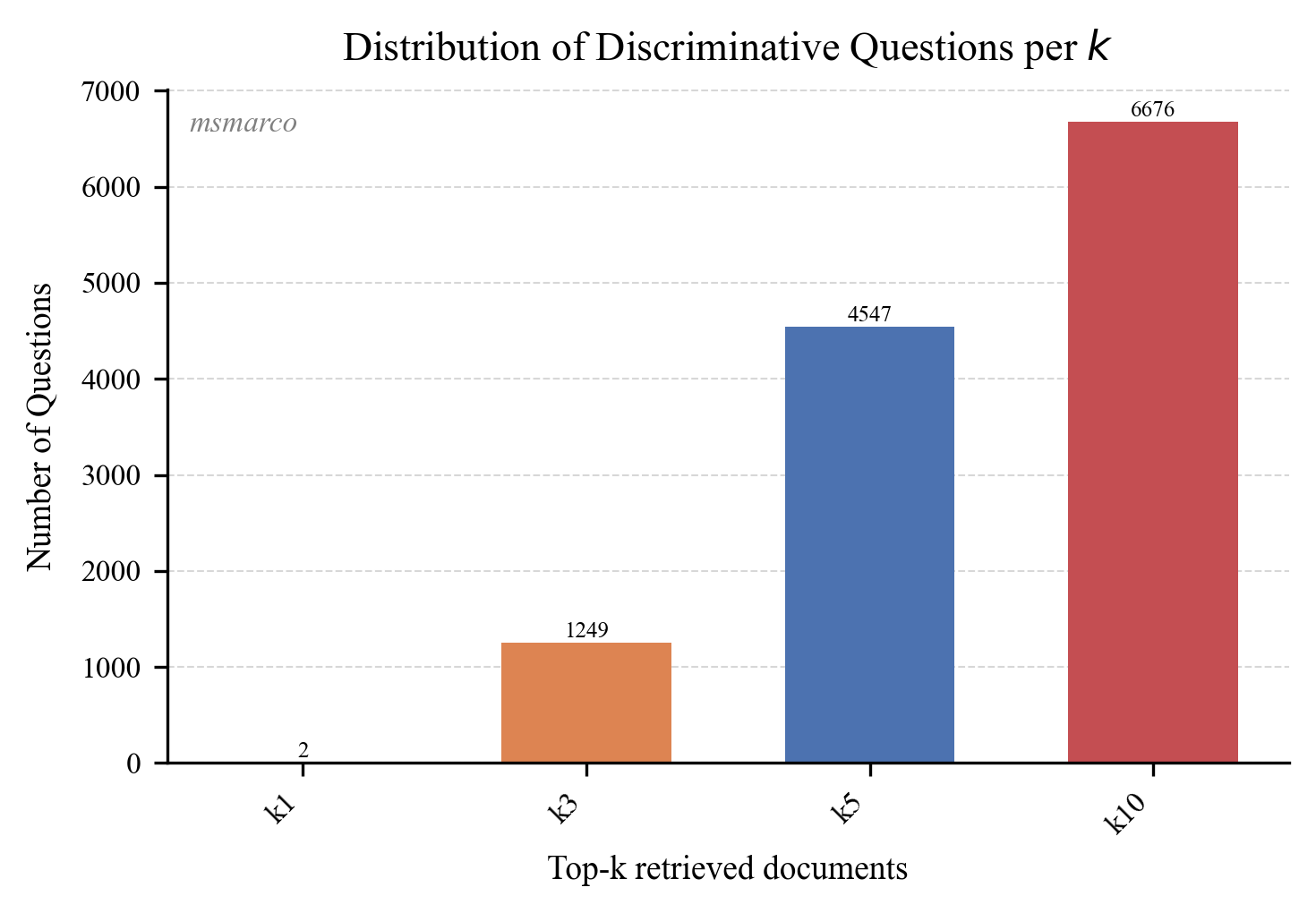}
    \caption{Effect of top-$k$ on the performance of the attack.}
    \label{fig:effect of top k}
\end{figure}

\paragraph{Effect of top-$k$ retrieved documents.}
The number of discriminative queries is directly influenced by the number of retrieved documents $k$.  
As shown in Figure~\ref{fig:effect of top k}, decreasing $k$ degrades the attack's performance as a restricted search space limits the opportunity to retrieve divergent documents across different models. 
Conversely, increasing $k$ improves the inference performance as an expanded retrieval pool naturally amplifies the probability that different models will retrieve distinct document sets.
These results are consistent with that ~\citet{beyondbench}, who demonstrated that retrieval similarity decreases with increasing $k$ for low values of $k$.

\paragraph{Effect of Reranker.} From an attacker’s perspective, the presence and architecture of a reranker in the target system are typically unknown. 
Nevertheless, EIA remains effective even in such situation. 
More precisely as shown in Table~\ref{table:discriminative_query_summary_k3}, when the target system uses a reranker but the candidate models do not, the EIA performance's improve for all candidate models when $k$ = 3. 
Furthermore, as previously explained, increasing the number of retrieved documents enhances the attack's performance by increasing the probability of models retrieving divergent document sets.

\paragraph{Adversarial discriminative queries.}
We evaluate three types of adversarial discriminative queries as described in Section~\ref{sec:adversarial_discriminative_queries} : (1) Random String denoted as Rnd. Str., (2) Non-queries denoted as Non-Q and (3) Probing Queries denoted as Prob.Q.
As shown in Table~\ref{table:discriminative_query_summary_k3}, all types of adversarial discriminative queries successfully discriminate the victim model from the candidate models.
Even when the target system employs a reranker, the attack remains effective but the number of successful adversarial discriminative queries decreases.

\subsection{Results on RAG}

To evaluate our attack in a realistic RAG environment, we conducted experiments with AnythingLLM using the MSMarco dataset. 
Unlike the controlled IR pipeline experiments, the attacker can only interact with the system through user queries and observe the retrieved document chunks returned by the framework. 
During the experimentation, we first observed that the query mode in AnythingLLM does not apply any retrieval threshold filtering unless we add a user-specified similarity threshold for the retrieval.
As a result, the retrieval behaviour is the same across different query categories both for query mode or chat mode.

\paragraph{Random strings and non-queries.}
For random string queries, the LLM used for the generation generally refused to produce answers as it often interprets the input as random characters or typographical noise. 
This behaviour was consistently observed across all evaluated LLMs ready-to-use in AnythingLLM, which includes (1) Qwen3 Vision 4B Instruct, (2) Gemma3 1B, (3) Llama3.2 3b, (4) pHI-3.5 3.8B and (5) Mistral 7b. %
Nonetheless, despite the refusal at the generation level, the retrieval system still returned top-$k$ document chunks indicating that retrieval was still performed even for  inputs that are semantically meaningful.  
This demonstrates that the retrieval component could remain accessible to the attacker even when the generation model rejects the query.

For the non-queries type, we observed different behaviors depending on whether the input contained a question mark. 
More precisely, when the non-query did not contain a question mark, the generation model typically responds with clarification or refusal messages such as ``there is no text content in any of the provided context''. 
In contrast, when a question mark was added to the same non-query, the generation model often produced generic confirmation responses such as ``yes'' or ``certainly''. %
Despite these differences in generated responses, the retrieval system still returned document chunks for all non-query inputs.

\paragraph{Probing queries.}
Probing queries successfully triggered both generation and retrieval behaviours, allowing the attacker to observe differences in the retrieved documents across embedding models.
The experiments also evaluated the impact of reranking with the AnythingLLM setup. 
We observed that adversarial discriminative queries remained effective when comparing an embedding model with reranking against different embedding models without reranking. 
The retrieved document chunks differed sufficiently to discriminate between candidate embedding models.

\begin{table}[h!]
    \centering
    \small
    \setlength{\tabcolsep}{1pt}
    \begin{tabular}{lcccc}
    \hline
    \textbf{Setup} & \textbf{Gen. Q.} & \textbf{Rnd. Str.} & \textbf{Non-Q.} & \textbf{Prob. Q.} \\
     & (max: 7433) & (max: 11) & (max: 11) & (max: 11) \\
    \hline
    IR              & 1249 & 11 & 11 & 11 \\
    IR + reranker   &  1860 & 9 & 7 & 11 \\
    RAG             &  ---  & 11 & 11 & 11 \\
    RAG + reranker  &  ---  & 11 & 11 & 11 \\
    \hline
    \end{tabular}
    \caption{Discriminative query results for $k=3$.}
    \label{table:discriminative_query_summary_k3}

    \begin{tabular}{lcccc}
    \hline
    \textbf{Setup} & \textbf{Gen. Q.} & \textbf{Rnd. Str.} & \textbf{Non-Q.} & \textbf{Prob. Q.} \\
     & (max: 7433) & (max: 11) & (max: 11) & (max: 11) \\
    \hline
    IR              & 4547 & 11 & 11 & 11 \\
    IR + reranker   &  4877 & 11 & 11 & 11 \\
    RAG             &  ---  & --- & --- & --- \\
    RAG + reranker  &  ---  & --- & --- & --- \\
    \hline
    \end{tabular}
    \caption{Discriminative query results for $k=5$.}
    \label{table:discriminative_query_summary_k5}
\end{table}

\section{Potential defense mechanisms}
\label{sec:defense}

Applying a similarity threshold reduces the effectiveness of the attack by filtering out retrieved documents that do not exceed the pre-defined threshold.
Our proposed defence is directly applied to the IR pipeline in which cosine similarity is determined between the query embedding and each retrieved document.
Documents whose similarity score falls below the predefined threshold are then discarded.
As shown in Table~\ref{tab:effect of threshold on EIA} for the generated queries, adding a minimum similarity threshold of 0.25 to the IR system had almost no impact as the system still produced 1247 discriminative queries for $k=3$. 
However, increasing the threshold to 0.6 significantly reduced the number of discriminative queries.

\begin{table}[h!]
    \centering
    \small
    \setlength{\tabcolsep}{3pt}
    \renewcommand{\arraystretch}{1.25}
    \begin{tabular}{lcccccccc}
    \toprule
    & \multicolumn{2}{c}{\textbf{Thr. 0.25}}
    & \multicolumn{2}{c}{\textbf{Thr. 0.50}}
    & \multicolumn{2}{c}{\textbf{Thr. 0.60}}
    & \multicolumn{2}{c}{\textbf{Thr. 0.75}} \\
    \cmidrule(lr){2-3}\cmidrule(lr){4-5}\cmidrule(lr){6-7}\cmidrule(lr){8-9}
    \textbf{Query}
      & $k{=}3$ & $k{=}5$
      & $k{=}3$ & $k{=}5$
      & $k{=}3$ & $k{=}5$
      & $k{=}3$ & $k{=}5$ \\
    \midrule
    Gen. Q.   & \textbf{1247} & \textbf{4545} & \textbf{580} & \textbf{2244} & \textbf{194} & \textbf{721} & \textbf{8} & \textbf{35} \\
    Rnd. Str. & \textbf{9} & \textbf{9} & \textbf{0} & \textbf{0} & \textbf{0} & \textbf{0} & \textbf{0} & \textbf{0} \\
    Non-Q.   & \textbf{10} & \textbf{10} & \textbf{0} & \textbf{0} & \textbf{0} & \textbf{0} & \textbf{0} & \textbf{0} \\
    Prob. Q.  & \textbf{11} & \textbf{11} & \textbf{0} & \textbf{0} & \textbf{0} & \textbf{0} & \textbf{0} & \textbf{0} \\
    \bottomrule
    \end{tabular}
    \caption{Effect of similarity retrieval threshold on the IR system.}
    \label{tab:effect of threshold on EIA}
\end{table}

While a similarity threshold can reduce the effectiveness of EIA, it may also negatively affect retrieval quality by filtering potentially relevant documents. 
To evaluate this trade-off, we measured retrieval performance on the MS MARCO passage retrieval benchmark using mean reciprocal rank at 10 and recall at 1000 (\textbf{MRR@10} and \textbf{Recall@1000}) with different similarity thresholds. 
\textbf{Docs/Retrieval} is simply how many documents are returned on average for each query, after applying the similarity threshold.

\begin{table*}[h!]
\centering
\small
\setlength{\tabcolsep}{4pt}
\begin{tabular}{llccccc}
\toprule
\textbf{Metric} & \textbf{Models} &
\textbf{0.00} & \textbf{0.25} & \textbf{0.50} &
\textbf{0.60} & \textbf{0.75} \\
\midrule

\multirow{5}{*}{MRR@10}
& e5-large-v2                           & 0.3587 & 0.3587 & 0.3587 & 0.3587 & 0.3587 \\
& gtr-t5-base                           & 0.3494 & 0.3494 & 0.3494 & 0.3494 & 0.3405 \\
& LaBSE                                 & 0.0603 & 0.0603 & 0.0462 & 0.0120 & 0.0005 \\
& all-MiniLM-L6-v2                      & 0.3046 & 0.3046 & 0.3045 & 0.2997 & 0.2092 \\
& paraphrase-multilingual-MiniLM-L12-v2 & 0.1957 & 0.1957 & 0.1957 & 0.1945 & 0.1460 \\

\midrule

\multirow{5}{*}{Recall@1000}
& e5-large-v2                           & 0.9777 & 0.9777 & 0.9777 & 0.9777 & 0.9777 \\
& gtr-t5-base                           & 0.9793 & 0.9793 & 0.9793 & 0.9793 & 0.7690 \\
& LaBSE                                 & 0.5419 & 0.5419 & 0.1456 & 0.0173 & 0.0007 \\
& all-MiniLM-L6-v2                      & 0.9653 & 0.9653 & 0.9316 & 0.8178 & 0.3937 \\
& paraphrase-multilingual-MiniLM-L12-v2 & 0.8586 & 0.8586 & 0.8421 & 0.7389 & 0.3397 \\

\midrule

\multirow{5}{*}{Docs/Retrieval}
& e5-large-v2                           & 1000.0 & 1000.0 & 1000.0 & 1000.0 & 998.6 \\
& gtr-t5-base                           & 1000.0 & 1000.0 & 1000.0 & 999.5 & 36.8 \\
& LaBSE                                 & 1000.0 & 999.7 & 47.3 & 1.0 & 0.0 \\
& all-MiniLM-L6-v2                      & 1000.0 & 1000.0 & 435.8 & 132.7 & 11.9 \\
& paraphrase-multilingual-MiniLM-L12-v2 & 1000.0 & 1000.0 & 835.2 & 357.3 & 29.7 \\

\bottomrule

\end{tabular}
\caption{Effect of similarity threshold on retrieval performance.}
\label{tab:threshold_performance}
\end{table*}

Table~\ref{tab:threshold_performance} shows that the impact of similarity threshold is model-dependent. 
For larger model such as \textit{e5-large-v2}, retrieval effectiveness remains the same even at a threshold of 0.75, with MRR@10 and Recall@1000 maintaining their original values. 
Similarly, \textit{gtr-t5-base} show small degradation up to a threshold of 0.60. 
However, \textit{LaBSE} is particularly sensitive to the similarity threshold as shown by the low score on MRR@10 and Recall@1000 dropping from 0.5419 to near zero.
For lighter models, such as \textit{all-MiniLM-L6-v2} and \textit{paraphrase-multilingual-MiniLM-L12-v2}, both MRR@10 and Recall@1000 scores decrease as the threshold increase.
These results suggest that even if similarity threshold can significantly reduce the attack, the choice of threshold should be adapted to the retrieval model on the system in order to preserve retrieval performance.

Another possible defence mechanism, for RAG system only, consists in hiding the retrieved document chunks from the user. 
This aligns with the notion of ``retrieved chunk leakage'' and ``answer leakage'' described in \citet{bodea2026sokprivacyrisksmitigations} in which the focus is on preventing sensitive content from being surfaced to end users.
While this limits the attacker’s ability to observe retrieved chunks and behaviour, it also reduces system transparency, as users can no longer verify the origin of the generated information.
\citet{Zhou2024TrustworthinessIR} survey paper discusses transparency as one of six key dimensions of trustworthiness in RAG systems, defined as: \textit{``Making RAG system processes and decisions clear and understandable to users, fostering trust and accountability''}.

Finally, one could introduce noise into the retrieved set by adding random or synthetic documents.
Although this may obscure the retrieval patterns exploited by the attacker, it might also negatively impact the quality and relevance of a RAG system’s responses.
\citet{cuconasu2024rethinking} shows that random documents and semantically related but non‑relevant document that do not contain the answer (called distractors) can possibly alter model behaviour risking reduced answer quality and accuracy.

\section{Conclusion}
\label{sec:conclusion}

In this work, we have proposed a novel attack called an embedding inference attack by which an adversary can infer the embedding model used by an information retrieval system.
This knowledge is often necessary to conduct embedding inversion attacks and can even unlock other privacy vulnerabilities.
Our primary motivation is the inversion of user queries, which may reveal private or sensitive information about the user.
The attack assumes access to the document collection used by the retrieval system, which is realistic as many IR applications rely on publicly available documents, as  well as access to a set of candidate embedding models, which is also realistic because many systems use open-source embedding models or publicly available APIs rather than embedding models developed in a private manner.

\section{Limitations}
\label{sec:limitation}

This work presents some limitations that could be addressed in future research. 
First, our evaluation does not investigate the impact of document chunk size and chunk overlap on the effectiveness of the attack.
Since these parameters directly influence the retrieval process in RAG systems, studying their effect could provide a better understanding of the robustness of different configurations.
In addition, our experiments mainly focus on open-source embedding models. 
Future work will extend the evaluation of the EIA to closed-source and proprietary systems, such as models provided by companies such as OpenAI or Google in order to assess whether the attack remains effective in such settings.

\bibliography{custom}

\end{document}